\documentclass[11pt,a4paper]{article}
\usepackage[margin=27mm,top=25mm,bottom=25mm]{geometry}
\usepackage{amsmath,amssymb,amsthm}
\usepackage{xcolor}
\usepackage[unicode,colorlinks=true,linkcolor=black,citecolor=blue!45!black,urlcolor=blue!45!black,bookmarksnumbered=true]{hyperref}
\hypersetup{pdftitle={Equilibrium Fluctuations and Exact Ties in Random Matrix Games},pdfauthor={Haokun Zhou},pdfsubject={Random matrix games, random convex hulls and matrix singularity}}
\numberwithin{equation}{section}
\newtheorem{theorem}{Theorem}[section]
\newtheorem{lemma}[theorem]{Lemma}
\newtheorem{proposition}[theorem]{Proposition}

\theoremstyle{definition}

\theoremstyle{remark}

\newcommand{\Pp}{\mathbb P}
\newcommand{\E}{\mathbb E}
\newcommand{\R}{\mathbb R}

\DeclareMathOperator{\Var}{Var}
\DeclareMathOperator{\sd}{sd}
\DeclareMathOperator{\supp}{supp}
\begin{document}
\begin{center}
{\LARGE\bfseries Equilibrium Fluctuations and Exact Ties\\[2pt] in Random Matrix Games\par}
\vspace{0.65em}
{\large Haokun Zhou\par}
\vspace{0.15em}
{\small Written in collaboration with Codex\par}
\end{center}
\begin{abstract}
How many actions does a player use in a large random zero-sum game, and how much does optimal play change the value? For an $n\times n$ game with independent entries from any fixed symmetric law of finite positive variance $\sigma^2$, the two support sizes have the same Gaussian fluctuation about $n/2$, with variance asymptotic to $n/8$. The centered value has an independent Gaussian fluctuation on scale $\pi\sigma/(2n)$. Every fixed positive absolute moment converges on this scale. We derive the two normals from the sum and difference of the probabilities that fresh actions satisfy the equilibrium inequalities. The geometry of these inequalities also gives the limiting distribution of strategy weights and exponential bounds for nonuniqueness, including for discrete payoffs. At finite size an exact sign average identifies the positive value with a random convex-hull distance. A rank identity then relates the exponential probability of an exact tie to the singularity probability of a random matrix. The resulting tie bases include $(3+2\sqrt2)/8$ for fair signs and $9/16$ for a four-point law.
\end{abstract}

\section{The game and its three observables}\label{sec:main}
Faris and Maier simulated games in which each player chose a probability distribution over $n$ actions after seeing the payoff matrix. Roughly half the actions went unused. For Gaussian and uniform payoffs of variance one, they estimated standard deviations of about $0.35\sqrt n$ for the number of unused actions and about $1.6/n$ for the value \cite[pp.~242--243]{FM}. The numbers suggest two different questions. Why is the support less variable than a set obtained by tossing $n$ independent fair coins? And why should choosing both strategies optimally leave a payoff of order $1/n$?

Jonasson proved that the mean support in the Gaussian game is asymptotic to $n/2$ \cite{Jon}. Berg and Engel studied the distribution of equilibrium weights by statistical-mechanical methods \cite{BE}. Cosson and Massouli\'e established the $1/n$ order of the Gaussian value standard deviation \cite{CM}. The joint law below gives the two numerical constants, their dependence on the entry law, and the dependence between the observables. The factor of two concerns the comparison with the \emph{binomial variance} $n/4$. Faris and Maier themselves recorded a narrower distribution.

Write $\Delta_d=\{x\in[0,\infty)^d:\sum_i x_i=1\}$. On a real $r\times c$ matrix $A$, the row player maximizes and the column player minimizes. The minimax theorem gives
\[
 V(A)=\max_{p\in\Delta_r}\min_{q\in\Delta_c}p^\top A q
     =\min_{q\in\Delta_c}\max_{p\in\Delta_r}p^\top A q .
\]
For any optimal $p,q$, the inequalities $Aq\le V(A)\mathbf 1$ and $p^\top A\ge V(A)\mathbf 1^\top$ hold coordinatewise. A positive coordinate of $p$ forces equality in the corresponding row inequality. A positive coordinate of $q$ does the same for a column inequality. Indeed, the weighted average of each set of inequalities is the equality $p^\top Aq=V(A)$. These are the active constraints that enter each of the arguments below.

Let $X$ have a fixed law symmetric about $m$, so $X-m$ and $m-X$ have the same distribution. Write $Y=X-m$. Let $A_n$ have $n^2$ independent entries distributed as $X$, let $B_n=A_n-m\mathbf1\mathbf1^\top$, and write $V_n=V(A_n)$ and $W_n=V(B_n)=V_n-m$. Subtracting $m$ shifts every mixed payoff by the same amount, so it preserves both optimal faces. An optimal face is the set of all optimal probability vectors for one player. Let $K_R,K_C$ denote the numbers of strictly positive coordinates of optimal row and column vectors. When there are several optima, take any measurable choices of attained support sizes on the realized matrix.

\begin{theorem}[Equilibrium on the original board]\label{thm:main}
Suppose $0<\sigma^2=\E Y^2<\infty$. For independent standard normal random variables $Z_1,Z_2$,
\begin{equation}\label{eq:main}
 \left(\frac{K_R-n/2}{\sqrt{n/8}},\
       \frac{K_C-n/2}{\sqrt{n/8}},\
       \frac{nW_n}{\sigma}\right)
 \ \Longrightarrow\ \left(Z_1,Z_1,\frac{\pi}{2}Z_2\right).
\end{equation}
For either support size $K$ and every fixed $s>0$,
\begin{equation}\label{eq:mom-main}
 \frac{\Var(K)}n\to\frac18,\qquad
 \frac{n\,\sd(V_n)}{\sigma}\to\frac{\pi}{2},\qquad
 \E\left|\frac{nW_n}{\sigma}\right|^s
       \to \left(\frac{\pi}{2}\right)^s\E|Z_2|^s .
\end{equation}
With probability at least $1-Ce^{-cn}$ both players have unique optimal strategies, and the two strategies have equal support size. The positive constants $c,C$ may depend on the law.
\end{theorem}

Two concrete continuous examples are a bounded symmetric law with a density and a bounded symmetric variable plus independent Gaussian noise of positive variance. The latter includes smoothed discrete inputs.

The same copy of $Z_1$ occurs twice in \eqref{eq:main}. The players' support fluctuations coincide in the limit, and their common limit is independent of the signed limiting value. There is a further distinction between using an action and using it heavily. About $n/2$ actions receive positive weight, but the sum of squared weights is asymptotic in probability to $\pi/n$. A uniform strategy on about $n/\pi$ actions would have the same sum.

Gaussian and fair-sign entries have these same limits. At finite size, fair signs can give an exact tie, as in the two-by-two matching-pennies game. To see what governs this event, let $T_n=\Pp(W_n=0)$ and $S_n=\Pp(\det B_n=0)$. For any nondegenerate symmetric law with an atom, if $\limsup S_n^{1/n}=b$, then $\limsup T_n^{1/n}=(1+\sqrt b)^2/4$ (Theorem~\ref{thm:tie-transfer}). Thus the same equilibrium has a universal fluctuation on the $1/n$ scale and a law-dependent probability of landing exactly at the center.

The proof follows three uses of the equilibrium inequalities. First we condition on a candidate support and ask how many unused actions pass its tests. The two resulting scores determine support and value (Section~\ref{sec:fluctuations}). Next we remove an action before testing whether it ties an optimum. This makes the tested action independent of the candidate optima and gives exact uniqueness (Section~\ref{sec:unique}). Section~\ref{sec:moments} proves the moment assertions by controlling unusually large payoffs. Finally, averaging the orientations of the inequalities converts a tie into a rank event (Section~\ref{sec:ties}). Throughout, the longer analytic inputs are stated where they are used and linked to their detailed proofs in the accompanying formal development \cite{Project}.

\section{From the active submatrix to the joint limit}\label{sec:fluctuations}

There are two sources of randomness in the support of an equilibrium. The entries on its active rows and columns determine the candidate strategies. The remaining rows and columns decide whether those candidates survive. We first separate these two experiments for an atomless entry law. We work with the centered board throughout, so its value is \(W_n\).

\subsection{Counting an equilibrium by its support}

Consider an \(r\times c\) board with \(r,c\ge1\). Atomless independent entries put it in general position almost surely. Indeed, the exceptional sets for the finitely many candidate bases are zero sets of nonzero polynomials. Such a set has product measure zero by conditioning on one variable at a time. In general position the two optimal strategies are unique and have the same support size \(K\). Fix \(k\) row labels and \(k\) column labels. If the resulting square subgame has a fully mixed equilibrium \((p,q)\) of value \(v\), let
\[
 F=\Pp(a^\top q\le v),\qquad G=\Pp(p^\top b\ge v),
\]
where \(a,b\) are independent fresh payoff vectors. The probabilities here hold \(p,q,v\) fixed. Write \(\nu_k\) for the law of the mark \(\theta=(F,G,v)\) restricted to the fully mixed event. Its total mass can be less than one. The joint distribution on the full board is
\begin{equation}
 \Pp_{r,c}(K=k,\,\theta\in d\theta)
 =\binom rk\binom ck F^{r-k}G^{c-k}\,\nu_k(d\theta).
 \label{eq:paper-mark}
\end{equation}
In fact, each unused row must earn at most \(v\) against \(q\), and each unused column must yield at least \(v\) against \(p\). These tests use disjoint sets of entries. The intersection of unused rows with unused columns can take any values. Summing over the support labels counts each equilibrium once.

One consequence will serve twice. Let \(Q_R=1-K/r\) and \(Q_C=1-K/c\). For \(0<h\le1\),
\begin{equation}
 \Pp(|Q_R-F|\ge h)\le 2e^{1/4-rh^2/16},\qquad
 \Pp(|Q_C-G|\ge h)\le 2e^{1/4-ch^2/16}.
 \label{eq:paper-pass}
\end{equation}
For the first bound, compare the weight in \eqref{eq:paper-mark} at \(r\) with its weights at \(r-d\) and \(r+d\). With \(d=\lfloor rh/4\rfloor\), the two resulting likelihood factors on the \(r\)-row board are
\[
 \frac{(r-K)_d}{(r)_d F^d},
 \qquad
 \prod_{j=1}^d\frac{F}{(r-K+j)/(r+j)},
\]
where \((a)_d\) is a falling factorial. Here \(F>0\) almost surely, and summing the corresponding weights bounds each expectation by one. If \(Q_R\ge F+h\), each factor of the first product is at least \(1+h/2\). If \(F\ge Q_R+h\), the same holds for the second product. Markov's inequality and \(\log(1+h/2)\ge h/4\) give the stated bound. The column calculation follows by interchanging the players and negating the payoffs. Symmetry and atomlessness place \(1/2\) between \(F\) and \(G\), since the two projection thresholds have opposite signs after reflection. On a square \(Q_R=Q_C\), so
\begin{equation}\label{eq:support-tail}
 \Pp\bigl(|K/n-1/2|\ge h\bigr)\le 4e^{1/4-nh^2/16}.
\end{equation}
This estimate for support uses no entry moments.

\subsection{Conditioning a random board on its dimensions}

Independently draw the two dimensions \(R,C\) from the Poisson law of mean \(n\). Work conditionally on \(R,C>0\), an event of probability \((1-e^{-n})^2\). Given the core size \(k\) and its mark, \(R-k,C-k\) are independent Poisson variables of means \(nF,nG\). To see this, multiply \eqref{eq:paper-mark} by the probabilities of the two dimensions and use, once on each side,
\[
 e^{-n}\frac{n^{k+u}}{(k+u)!}\binom{k+u}{k}F^u
 =\frac{n^k e^{-n(1-F)}}{k!}\,
   \frac{e^{-nF}(nF)^u}{u!}.
\]
The same identity factors over any finite collection of independent Poisson counts when a candidate has that product of exterior pass probabilities.

Introduce the two coordinates
\[
 U_1=\frac{k+nF-n}{\sqrt n},\qquad U_2=\frac{k+nG-n}{\sqrt n}.
\]
The centered dimension vector divided by \(\sqrt n\) is \(U\) plus a conditionally centered Poisson fluctuation. Since \(R/n,C/n\to1\), \eqref{eq:paper-pass} and the fact that \(F,G\) straddle \(1/2\) imply \(F,G\to1/2\) in probability. At each fixed frequency \(t=(t_1,t_2)\), the conditional characteristic function of the latter fluctuation therefore converges in \(L^1\) to \(q(t)=\exp(-\|t\|^2/4)\). For example, its first factor is
\(\exp\{nF(e^{it_1/\sqrt n}-1-it_1/\sqrt n)\}\).
The complete dimension vector has limiting characteristic function \(\exp(-\|t\|^2/2)\). If \(q_n(t)\) denotes the conditional noise characteristic function, then
\[
 \bigl|\phi_{\rm dim}(t)-q(t)\phi_U(t)\bigr|
 \le \E|q_n(t)-q(t)|.
\]
Dividing by \(q(t)>0\) gives \(U\Rightarrow N(0,I_2/2)\) in the random-size experiment. This argument only requires conditional control of the noise. The Lean game proof obtains the same limit by the bilateral Laplace transform.

The fixed-square experiment puts a second Gaussian factor in this calculation. Write \(p_\lambda(j)\) for the Poisson mass, set it to zero for negative \(j\), and put
\[
 L_n=\frac{p_{nF}(n-k)p_{nG}(n-k)}{p_n(n)^2},
 \qquad H(u)=2e^{-\|u\|^2}.
\]
The density of the core conditioned on \(R=C=n\), relative to its positive-dimension law, is \((1-e^{-n})^2 L_n\). On a bounded score window, the straddling property forces \(n-k=n/2+O(\sqrt n)\) and \(nF,nG=n/2+O(\sqrt n)\). The Poisson local limit obtained by Stirling's formula gives \(L_n-H(U)\to0\) uniformly there. Off that window \(L_n\ge0\) and \(H\) is uniformly small. Consequently \(\E(H(U)-L_n)_+\to0\). Also \(\E L_n\to1\) by normalization, and \(\E H(U)\to1\) by the mixed Gaussian limit. These two identities give \(\E|L_n-H(U)|\to0\). Multiplying the mixed density \(\pi^{-1}e^{-\|u\|^2}\) by \(H(u)\) now proves
\begin{equation}
 U\ \Longrightarrow\ N(0,I_2/4)\qquad\text{on the fixed square}.
 \label{eq:paper-score}
\end{equation}
This use of the total mass permits large likelihoods on rare cores.

\subsection{The geometry of a fresh payoff}

The value enters through \(F-G\). To recover it at order \(1/n\), one needs the mass in a short projection interval. The length of the interval depends on the norm of the actual equilibrium strategy. The following two estimates supply both pieces. Write \(Z_+=\max(Z,0)\) for \(Z\sim N(0,1)\), let \(\nu=\mathcal L(\sqrt2 Z_+)\), and, for a nonnegative unit vector \(x\), set \(\nu_x=n^{-1}\sum_j\delta_{\sqrt n x_j}\). The quadratic Wasserstein distance \(W_2\) is the infimum of the root mean square distance over couplings.

\begin{lemma}[Direction exclusion and projection intervals]\label{lem:paper-geometry}
Let the entries be centered, symmetric and of variance \(0<\sigma^2<\infty\).
For each \(\epsilon>0\) there are \(b,c,C>0\) such that
\[
 \Pp\{\exists\,x\ge0:\ \|x\|_2=1,\ W_2(\nu_x,\nu)\ge\epsilon,\
                  \max_i (B_nx)_i\le b\}\le Ce^{-cn}.
\]
If \(W_2(\nu_{x_n},\nu)\to0\) for deterministic nonnegative unit vectors \(x_n\), let
\(S_n=\sum_j\sqrt n\,x_{nj}Y_j\) for fresh independent copies of \(Y=X-m\). For each fixed \(a>0\),
\[
 \sup_{z\in\R}\left|\sqrt n\,\Pp\{S_n\in[z,z+a]\}
       -a\,\varphi_\sigma(z/\sqrt n)\right|\longrightarrow0,
\]
where \(\varphi_\sigma\) is the centered normal density of variance \(\sigma^2\).
\end{lemma}

We give the main estimates behind the lemma. The full geometric comparison and the Fourier approximation are proved in \href{https://github.com/Sheltercosmo/minimax-universality/blob/557c5e4321a3d9d2dcf969999462e8f380f8aa05/proofs/matrix-games/MatrixGames/FiniteVarianceDirectionExclusion.lean}{the direction-exclusion theorem} and \href{https://github.com/Sheltercosmo/minimax-universality/blob/557c5e4321a3d9d2dcf969999462e8f380f8aa05/proofs/matrix-games/MatrixGames/ProjectionIntervalAsymptotics.lean}{the projection interval theorem}. For the first assertion standardize the variance to one and truncate the entries, renormalizing afterwards. Put coordinates of \(x\) larger than \(n^{-2/5}\) in a set \(H\), so \(|H|\le n^{4/5}\). Retain those columns and replace the remaining entries by Gaussians in a smoothed minimax problem. The test vectors \(y\ge0\) have \(\|y\|_2\le1\), \(\|y\|_\infty\le L/\sqrt n\). Thus each replaced entry has coefficient at most \(L n^{-9/10}\). Taylor replacement cancels its first two moments. With smoothing parameter of order \(\sqrt n\), the sum of the third-order errors is \(O(n^{3/10})=o(\sqrt n)\) for a fixed truncation and fixed mesh accuracy.

Write \(x=(z,w)\) on \(H,H^c\). Gordon's comparison \cite{Gordon} conditions on the retained matrix \(B_H\), which appears on both sides of the inequality. Integrate this comparison over \(B_H\). Under the resulting joint law of \(B_H\) and an independent standard Gaussian vector \(g\), the auxiliary row vector is \(\|w\|_2g+B_Hz\). For each fixed unit \(x\), its coordinates are independent, symmetric and of variance one. Their positive parts therefore have second moment \(1/2\). Uniform integrability of the squared linear combinations permits one clipping level for all sufficiently large truncations. For each \(H\), concentration of the clipped row energies on a net of the at most \(1+|H|\) parameters \((\|w\|_2,z)\) gives a lower bound arbitrarily close to \(\sqrt{n/2}\) for their Euclidean norm.

The other Gaussian term is \(h^\top w\), with \(h\) standard Gaussian. Its most negative unit direction is the normalized negative part of \(h\). After padding by zeros on \(H\), its empirical law in unit-vector coordinates approaches \(\nu\). At a fixed distance from this law, the equality condition in Cauchy--Schwarz gives a positive gap of order \(\sqrt n\) in the auxiliary minimax bound. Finite nets for the two vector sets and the subexponential number of possible \(H\) preserve the gap. \href{https://github.com/Sheltercosmo/minimax-universality/blob/557c5e4321a3d9d2dcf969999462e8f380f8aa05/proofs/matrix-games/MatrixGames/HybridDirectionExclusion.lean}{The heavy-column comparison} gives the uniform estimates.

To return to the original entries let \(E\) be their centered residual matrix, of entry standard deviation \(\tau\). Row symmetrization and contraction give \(\E\|E\|_{2\to1}\le 3n\tau\) for a square. Thus the expected change of the average clipped positive row payoff, uniformly in unit directions, is \(O(\tau)\). Fix the truncation so this costs less than the gap. Once the clipping level is fixed, changing one entire row changes the clipped average by \(O(1/n)\). Bounded differences give the exponential probability estimate.

For the interval assertion, the squared tails of the empirical coefficients \(b_j=\sqrt n\,x_{nj}\) are uniformly integrable, hence \(\max_j |b_j|/\sqrt n\to0\). The low-frequency product \(\prod_j \E e^{iu b_jY_j/\sqrt n}\) has Gaussian limit by the Lindeberg estimate. On each fixed frequency band away from zero it is the product at the unscaled frequencies, \(\prod_j|\E e^{it b_jY_j}|\), that decays exponentially. Indeed, the set \(\{v:|\E e^{ivY}|=1\}\) is a proper closed subgroup of \(\R\), whereas the positive part of \(\nu\) has a density on every positive interval. Integrating \(1-|\E e^{itbY}|\) against a compactly supported cutoff there gives a strictly positive lower bound uniformly on the band. Near zero, the coefficients in the same interval give a bound \(e^{-cnt^2}\). Fourier inversion against band-limited upper and lower approximations of the interval indicator proves the assertion. A translation contributes a phase of modulus one, which gives uniformity in \(z\). In particular even an arithmetic entry law has the stated interval limit along these directions.

\subsection{Inverting the scores and removing the perturbation}

On a board with \(W_n\le0\), every optimal column vector \(w\) satisfies \(B_n w\le0\). On \(W_n\ge0\), every optimal row vector satisfies the corresponding inequalities for \(-B_n^\top\). Lemma~\ref{lem:paper-geometry} therefore gives the target profile for the unit direction \(x=w/\|w\|_2\) on one side at once. Since
\[
 \int u\,d\nu_x(u)=\frac{1}{\sqrt n\|w\|_2},
 \qquad \int u\,d\nu(u)=\frac{1}{\sqrt\pi},
\]
the norm on that side is asymptotic in probability to \(\sqrt{\pi/n}\).

On an atomless board the relevant fresh-action probability differs from \(1/2\) by the projection mass between zero and \(s=\sqrt n|W_n|/\|w\|_2\). For directions sufficiently close to \(\nu\), every interval of length one starting in \([0,\sqrt n]\) has probability at least \(c/\sqrt n\) for large \(n\), by the uniform interval estimate and compactness in the normalized center. Summing these intervals gives a bound below by \(c\lfloor\min(s,\sqrt n)\rfloor/\sqrt n\). Equation~\eqref{eq:paper-pass} and the straddling property now give tightness of \(nW_n\). They also give \(\Pp(\sqrt n|W_n|>\delta)\le C_\delta e^{-c_\delta n}\) for each \(\delta>0\). Since every simplex strategy has norm at least \(1/\sqrt n\), its normalized optimal direction then has all constraint payoffs at most \(\sqrt n|W_n|\), on the appropriate orientation of the board. Applying the exclusion estimate a second time gives the unit profile for \emph{both} optimal faces.

At bounded signed thresholds the local interval estimate, symmetry and monotonicity give
\(\sqrt n[\Pp(S_n\le s)-1/2]=s\varphi_\sigma(0)+o(1)\), uniformly on compact sets. Apply this first with \(s=\sqrt nW_n/\|q\|_2\), then with \(s=-\sqrt nW_n/\|p\|_2\). The fresh vectors are independent of the board. Tightness of \(nW_n\) and the two norm estimates give
\[
 \sqrt n(F-1/2)=\frac{nW_n}{\pi\sqrt2\,\sigma}+o_{\Pp}(1),\qquad
 \sqrt n(G-1/2)=-\frac{nW_n}{\pi\sqrt2\,\sigma}+o_{\Pp}(1).
\]
Adding and subtracting the definitions of \(U_1,U_2\) yields
\[
 \frac{K-n/2}{\sqrt n}=\frac{U_1+U_2}{2}+o_{\Pp}(1),\qquad
 \frac{nW_n}{\sigma}=\frac{\pi}{\sqrt2}(U_1-U_2)+o_{\Pp}(1).
\]
Under \eqref{eq:paper-score} the half-sum has variance \(1/8\), the rescaled difference has variance \(\pi^2/4\), and their covariance is zero. Joint Gaussianity gives independence.

For an arbitrary symmetric law add independent bounded symmetric atomless noise of amplitude \(\eta_n=o(1/n)\) to every cell. The formal construction uses \(\tfrac12\arctan G\) for a standard normal \(G\). The atomless argument holds along this coupled sequence because the entry residual tends to zero in \(L^2\), the geometry transfers by the residual estimate, and the projection characteristic functions converge uniformly on compact bands. The value changes by at most \(\eta_n\) deterministically. In particular the fixed-tolerance bound for \(\sqrt n|W_n|\) and the two-sided direction exclusion apply to the original board as well.

To recover support on an atomic board, we must control which tight actions can disappear as the perturbation vanishes. For any optimizer \(w\), write \(s(w)\) for its number of positive entries and \(t(w)\) for the number of opposing pure actions tight against it. The finite-variance surplus estimate gives a law-dependent integer \(d\) such that, outside probability \(Ce^{-cn}\), \(t(w)\le s(w)+d\) for every optimizer on both faces. This estimate is proved by screening possible excess constraints and excluding optimal vertices with excessive fresh-row point mass.

Here is the deterministic step that makes this estimate sufficient. Suppose regular boards with common support \(k\) approach \(B_n\). Compactness, passage to a subsequence with fixed labels, and continuity of the saddle inequalities give optimal vectors \(p_0,q_0\) of support at most \(k\), each with at least \(k\) tight opposing actions. For any optimal row vector \(p\), complementary slackness gives
\[
  s(p)\le t(q_0)\le k+d,\qquad
  s(p)+d\ge t(p)\ge s(q_0)\ge t(q_0)-d\ge k-d .
\]
The column argument is identical. Thus \(k-2d\le s(w)\le k+d\) for every original optimizer. At fixed \(n\) there are only finitely many candidate sizes, and the sets admitting the limiting witnesses are closed. A sufficiently small perturbation therefore has its support size in the envelope of the original board. Choosing its deterministic amplitude sufficiently small in each dimension makes the probability of an envelope failure at most \(e^{-n}\). Together with the surplus estimate, this gives discrepancy at most \(2d\) outside an event of probability at most \(Ce^{-cn}+e^{-n}\). Dividing by \(\sqrt n\) proves the duplicated support coordinate for the two selections on the same board. The squared normalized discrepancy also tends to zero in expectation, since it is \(O(1/n)\) off the exceptional set and \(O(n)\) on it.

Return to the probability weights by writing \(nw_j=(\sqrt n\|w\|_2)(\sqrt n x_j)\). For both optimal faces simultaneously,
\[
 W_2\!\left(\frac1n\sum_j\delta_{nw_j},\,\mathcal L(\sqrt{2\pi}Z_+)\right)
 \ \longrightarrow\ 0,\qquad n\sum_j w_j^2\ \longrightarrow\ \pi
\]
in probability, with an exponentially small failure probability at each fixed tolerance. Thus the effective number \(1/\sum_j w_j^2\) is asymptotic to \(n/\pi\). The fraction of positive coordinates tends to \(1/2\) uniformly over the faces by the support bounds. For a bounded Lipschitz test \(f\) with \(f(0)=0\), the positive-coordinate mean is the all-coordinate mean divided by this fraction. It converges to \(\E f(\sqrt{2\pi}|Z|)\), uniformly for bounded \(1\)-Lipschitz tests of sup norm at most one. Subtracting \(f(0)\) gives the usual bounded Lipschitz convergence. The support count is needed here because the limiting law on all coordinates has an atom at zero.

\section{When is the optimal strategy unique?}\label{sec:unique}

The value varies continuously with the entries of the matrix. The support
of an optimal strategy can jump under an arbitrarily small perturbation.
For example, consider the skew-symmetric game
\[
 A(t)=
 \begin{pmatrix}
 0&t&-1&2\\ -t&0&2&-1\\ 1&-2&0&0\\ -2&1&0&0
 \end{pmatrix}.
\]
Its value is zero. At $t=0$ the optimal face of either player consists of
$(s,1-s,0,0)$ with $1/3\le s\le2/3$. For every $t>0$, both players have
the unique optimum $(2,1,t,0)/(3+t)$. To see the change directly, write
$A(t)q\le0$ for an optimal column strategy. The third row gives
$q_1\le2q_2$. Twice the first inequality plus the second gives
$t(2q_2-q_1)+3q_4\le0$. Thus $q_4=0$ and, if $t>0$, $q_1=2q_2$.
The first two rows then give $q_3=tq_2$. At $t=0$ those same rows give
$q_3=q_4=0$, and the last two give the stated bounds on $s=q_1$.
Skew-symmetry identifies the two optimal faces. The example explains why a
proof about support on an atomic board must control that board itself.

\subsection{An extra tight constraint witnesses nonuniqueness}

Write $v$ for the value of a fixed matrix $B$. For a column optimum $q$,
let $I(q)=\{i:(Bq)_i=v\}$. For a row optimum $p$, put
$J(p)=\{j:(p^\top B)_j=v\}$. A tight action can have zero weight.
Complementary slackness nevertheless gives
\begin{equation}\label{eq:unique-inclusions}
 \supp p\subseteq I(q),\qquad \supp q\subseteq J(p)
\end{equation}
for \emph{every} optimal pair. Indeed, a convex combination of numbers
bounded above by $v$ can equal $v$ only if every term of positive weight
equals $v$. The lower bound for the row player gives the second inclusion.

\begin{lemma}\label{lem:unique-vertices}
Suppose $p,q$ are vertices of their respective optimal faces. If
$|I(q)|\le|\supp q|$ and $|J(p)|\le|\supp p|$, then both faces are
singletons, and $|\supp p|=|\supp q|$.
\end{lemma}
\begin{proof}
Chain the two assumed inequalities with \eqref{eq:unique-inclusions}.
All four cardinalities are equal, so $\supp p=I(q)$ and $\supp q=J(p)$.
For any other column optimum $q'$, \eqref{eq:unique-inclusions} now says
that $q'$ vanishes wherever $q$ vanishes and is tight at every row tight
for $q$. If $q'\ne q$, all those equalities persist along
$q+\varepsilon(q-q')$. Every remaining inequality is strict at $q$,
so it remains feasible for all sufficiently small $\varepsilon>0$.
This puts $q$ inside a nontrivial feasible segment, contradicting
extremality. The row argument follows by applying the same proof to
$-B^\top$.
\end{proof}

In particular, failure of uniqueness or of equal support sizes yields a
vertex on at least one side with more tight opponent actions than positive
coordinates. Such a surplus gives us a constraint to hold out.

\begin{theorem}[Exponential uniqueness]\label{thm:paper-uniqueness}
For a fixed symmetric entry law of finite positive variance there are
$C,c>0$ such that for every $n\ge1$, with probability at least
$1-Ce^{-cn}$, each player has a unique optimal strategy and the two
supports have the same cardinality. The same bound holds without a moment
assumption if the symmetric law has a positive diffuse component.
\end{theorem}

\begin{proof}
Subtract the center from every entry. This preserves both optimal faces.
Consider a column vertex $q$ with support $J$ of size $k$ and
$|I(q)|>k$. There is a set $I\subset I(q)$ of $k$ rows for which the
bordered system
\begin{equation}\label{eq:unique-bordered}
 B_{I,J}q_J=v\mathbf 1,\qquad \mathbf 1^\top q_J=1
\end{equation}
determines $(q_J,v)$ uniquely. To see this, first include \emph{all}
tight rows and consider a homogeneous solution $(d,s)$ supported on
$J$, with $\mathbf 1^\top d=0$. Multiplication of the tight equations
by the weights of an optimal $p$ gives $s=p^\top Bd=0$, since
$p$ is supported on $I(q)$ and $p^\top B$ is constant on $J$.
A nonzero $d$ would allow both $q+\varepsilon d$ and $q-\varepsilon d$
as optima for small $\varepsilon>0$. Thus the homogeneous system has
trivial kernel. Extend the mass equation to a basis of the row space
using $k$ tight equations. Choose any $i\in I(q)\setminus I$.
The pair in \eqref{eq:unique-bordered} uses no entry of row $i$,
yet the omitted row must satisfy $B_iq=v$ exactly.

The choice of $(I,J)$ can still depend on row $i$. We use the following
two estimates from the formal development. They specify which probability
is small before any union over possible bases.

\begin{enumerate}
\item For some $e>0$, outside an event of probability at most $C_1e^{-c_1n}$,
every column vertex with at least $n/4$ positive coordinates satisfies
\[
  \sup_{y\in\R}\Pp_R\{R q=y\}\le e^{-en},
\]
where $R$ is an independent row and $Rq$ denotes its scalar product with
$q$. This estimate concerns optima simultaneously. Its proof normalizes
$q$ to a Euclidean unit vector and rules out near integer-lattice
resonances up to frequency $\exp(bn)$. For the remaining vectors the
small-interval estimate at radius $H\exp(-bn)$ is bounded by
$C(H\exp(-bn)+\exp(-c\kappa^2n))$ with positive law-dependent constants.
The bound holds at every center and hence for each singleton.

\item Given $\kappa_0>0$, the data $D_i$ consisting of all rows except $i$
determine a measurable list $\mathcal C_i(D_i)$ of at most
$\exp(\kappa_0 n)$ pairs of index sets. Outside an event of probability
at most $C_2e^{-c_2n}$, all column vertices have support at least $n/4$,
and the lists contain the pairs needed in \eqref{eq:unique-bordered}.
Here the preliminary bounded-surplus estimate supplies a fixed $D$ such
that $|I(q)|\le|\supp q|+D$ outside an exponentially rare event. The
construction enumerates feasible support labels after deletion, allowing
for the few large entries in a reinstated row. It enlarges the row-label
sets by a Hamming ball of radius $D$ to cover the choice of a basis.
The support localization and the binomial bound on the number of nearby
labels give the stated list size. A list exceeding the cap on exceptional
data may be replaced by the empty list.
\end{enumerate}

Take $\kappa_0=e/2$. For each fixed $D_i$ solve
\eqref{eq:unique-bordered} for every listed pair with $i\notin I$ and an
invertible system. Keep only those solutions $(q',v')$ for which an
independent row hits $Rq'=v'$ with probability at most $e^{-en}$.
These tests use only the retained data and the entry law. The two estimates
ensure that a surplus on a good full board produces a retained solution.
Conditional on $D_i$, the probability of any retained hit is at most
$e^{en/2}e^{-en}$. Integrating, then summing over $i$, bounds the
one-sided failure probability by
\[
 C_1e^{-c_1n}+C_2e^{-c_2n}+n e^{-en/2}.
\]
The law of $-B_n^\top$ is the law of $B_n$, so doubling this bound covers
both sides. Lemma~\ref{lem:unique-vertices} concludes the argument.
Enlarging the prefactor covers the finitely many small dimensions.

For the assertion without moments, let $\alpha<1$ be the total mass of
the atoms and fix $\alpha/2<a<1/2$. The mixed-law proof uses two different
estimates. With exponentially high probability, there are vertices on
both sides having at least $an$ tight opponent actions. The probability
of a surplus vertex with at least $an$ tight actions is exponentially
small. The latter estimate treats positive values by counting candidate
subgames, negative values by a conditional rectangular sign bound, and
zero values by the finite rank identity. Applied on both sides, these
estimates give the hypotheses of Lemma~\ref{lem:unique-vertices}.
\end{proof}

Let $L_n,H_n$ denote the minimum and maximum support sizes attained by
either player. Both are measurable. One can test each of the finitely
many possible support masks by feasibility of linear inequalities on
the optimal faces. Before the deletion argument, a surplus bound $D$ on
both faces already gave $H_n-L_n\le2D$ since an opposite-side pair differs
by at most $D$ by \eqref{eq:unique-inclusions}, and a common opponent
compares any same-side pair. Theorem~\ref{thm:paper-uniqueness} improves
this to $\Pp(H_n>L_n)\le Ce^{-cn}$. Even when $H_n=L_n$, several vectors
of that same support size can be optimal, as $A(0)$ illustrates.

\subsection{Finite-size obstructions}

\begin{proposition}\label{prop:finite-unique}
For any real i.i.d.\ entry law and fixed $n\ge2$, both players have
unique optima almost surely if and only if the entry law is atomless.
This is also equivalent to atomlessness of the game value. If
$p=\Pp(X=a)>0$, then
\begin{align}
 \Pp(\text{some player has multiple optima})&\ge p^{3n-2}, \label{eq:atom-cross}\\
 \Pp(V_n=a)&\ge p\,\Pp(X\le a)^{n-1}\Pp(X\ge a)^{n-1}. \label{eq:pure-atom}
\end{align}
\end{proposition}
\begin{proof}
For independent atomless entries, a nonzero polynomial in the entries
vanishes with probability zero. Induct on the number of variables,
conditioning on all but the last and applying the induction hypothesis
to a nonzero coefficient. For $k$ columns and $k+1$ rows, the determinant
of the $k$ row differences from one anchor row on these columns is a
nonzero polynomial, since the zero anchor and the rows of the identity
matrix give determinant one. Simultaneously for all choices of labels on both sides,
these determinants are nonzero. No vector supported on $k$ columns
can therefore be a probability vector giving the same payoff on $k+1$
rows. Lemma~\ref{lem:unique-vertices} gives uniqueness.

For each fixed $a$, translate all entries by $-a$. Every submatrix of
the translated board has full rectangular rank almost surely, by the
same polynomial argument. If the value were zero, the submatrix on an
optimal support pair would have a nonzero right null vector $q_J$ and a
nonzero left null vector $p_I$, by complementary slackness. This would
give rank strictly less than both dimensions. Thus $\Pp(V_n=a)=0$.

Conversely, prescribe one entire row and two entire columns to equal an
atom $a$. The row guarantees $a$ and either column guarantees $a$ for
the minimizer. The columns give distinct pure optima. Counting the
$3n-2$ prescribed entries gives \eqref{eq:atom-cross}.
A pure saddle at a specified cell instead needs that cell equal to $a$,
the rest of its row at least $a$, and the rest of its column at most $a$.
Independence gives \eqref{eq:pure-atom}. Both bounds are positive.
\end{proof}

For fair signs the endpoint events give a stronger bound. The value is
$1$ exactly when some row consists entirely of $+1$, since any row used
by a strategy guaranteeing $1$ must attain $1$ in every column. Independently
across rows, this has probability $1-(1-2^{-n})^n$. At value $1$ every
column strategy is optimal. The reflected event $V_n=-1$ has the same
probability and makes every row strategy optimal. The two events are
disjoint, giving the lower bound $2[1-(1-2^{-n})^n]$ when $n\ge2$.

One final witness explains the law dependence of the upper exponential
rate. Suppose $p_0=\Pp(Y=0)>0$ and the centered law is symmetric and
nondegenerate. Require the first row to vanish and the value of the
minor obtained by deleting that row and one prescribed column to be
strictly negative. The pure first row guarantees zero. Extend an optimal
minimizer of the minor by zero on the omitted column, giving strictly
negative payoff to every other row. Mixing a sufficiently small amount
of the uniform strategy into this vector preserves all strict inequalities
and yields a full-support column optimum. Complementary slackness forces
the first row to be the unique row optimum. Hence $L_n=1$ and $H_n=n$.
The two parts of the event use disjoint entries, so its probability is
\[
 p_0^n\Pp(W_{n-1}<0)
   =\tfrac12 p_0^n\bigl(1-\Pp(W_{n-1}=0)\bigr)
   =(\tfrac12+o(1))p_0^n .
\]
The last step uses tie decay for a nondegenerate symmetric game, which
requires no moments. For any fixed width allowance, a bound $Ce^{-cn}$
therefore requires $c\le-\log p_0$. As $p_0$ tends to one, no positive
exponent can serve all these laws.

\section{Moments and their sharp finite-size boundary}\label{sec:moments}
A limit in distribution describes bounded continuous tests. The variance of the support and the moments of the value require separate tail estimates. The two observables have different tails. The support is bounded by the dimension, whereas a single row of large payoffs can give an arbitrarily large value.

\subsection{What a finite board can integrate}
\begin{proposition}\label{prop:moment-boundary}
For $t>0$ let $\bar F(t)=\Pp(Y>t)$. On a square board,
\begin{equation}\label{eq:tailpower}
 \bar F(t)^n\le \Pp(W_n>t)
 \le [1-(1-\bar F(t))^n]^n\le n^n\bar F(t)^n .
\end{equation}
For a symmetric law and $s>0$, $\E|W_n|^s$ is finite exactly when
$\int_1^\infty t^{s-1}\bar F(t)^n\,dt$ is finite. A finite entry variance guarantees finiteness for $s\le 2n$, including the endpoint. There exists one symmetric finite-variance law for which $\E|W_n|^s=\infty$ for \emph{every} $n\ge1$ and $s>2n$.
\end{proposition}
\begin{proof}
If every entry of the first row exceeds $t$, the row player guarantees more than $t$. If a column has all entries at most $t$, the column player guarantees at most $t$. Independence of the columns and the union bound give \eqref{eq:tailpower}. The symmetry $-B_n^\top\stackrel d=B_n$ and the identity $V(-B_n^\top)=-W_n$ allow us to integrate twice the positive tail. This gives the criterion.

Set $M_j=\max_i (B_{ij})_+$. Then $(W_n)_+\le\min_j M_j$, so
\begin{equation}\label{eq:critical-integrability}
 \E |W_n|^{2n}\le 2\,\E\prod_{j=1}^n M_j^2
              \le 2\big(n\,\E Y_+^2\big)^n .
\end{equation}
This step retains the endpoint that a direct integration of a Chebyshev bound would lose. For sharpness, put mass $a_k=e^{-3-k-2k^2}$ at each of $\pm e^{k^2}$, for $k=0,1,\ldots$, and put the remaining mass at zero. Its second moment is $2e^{-3}\sum_k e^{-k}<\infty$. If the first row consists entirely of $e^{k^2}$ then $W_n\ge e^{k^2}$. These events are disjoint over $k$, whence
\[
 \E |W_n|^s \ \ge\ \sum_{k\ge0} \exp\big((s-2n)k^2-nk-3n\big).
\]
The series diverges whenever $s>2n$.
\end{proof}

\subsection{Why all fixed orders converge}
Work in units $\sigma=1$, and put $H_n=nW_n$. The score estimates from Section~\ref{sec:fluctuations} have the following quantitative consequence. For large $n$, there is a measurable exceptional event $D_n$ and positive law-dependent constants such that
\begin{equation}\label{eq:good-bad}
 \Pp(D_n)\le C e^{-cn},\qquad
 \Pp(|H_n|>t,\ D_n^c)\le C'e^{-c't^2}\quad(t>0).
\end{equation}
One can take $D_n=\{|H_n|>\sqrt n\}$. Here is how the cutoff enters. On the side that can secure a payoff of the winning sign, the normalized optimal direction has the profile of Lemma~\ref{lem:paper-geometry}. The mean of its scaled coordinates is close to $1/\sqrt\pi$. Consequently, raising the test level for $n|W_n|$ by one tests a projection interval of length bounded above and below by positive constants. Up to the cutoff $\sqrt n$, the center of this interval divided by $\sqrt n$ stays in a fixed compact set. The limiting Gaussian density is bounded below there. The uniform projection interval estimate now shows that every such unit increase changes the scaled passing probability by at least a fixed positive amount. Summing these increments and using the deletion bounds \eqref{eq:paper-pass} gives a Gaussian tail up to the cutoff, with an additive error $Ce^{-cn}$ from exceptional directions. At the cutoff both terms are exponentially small in $n$. Below it the same additive error is bounded by a Gaussian tail in $t$. Bounded smoothing and the value Lipschitz inequality carry this estimate to atomic laws.

Fix $s>0$ and choose $r>s$. For $n\ge2r$, Cauchy--Schwarz and \eqref{eq:critical-integrability} give
\[
 \E[|H_n|^r\mathbf1_{D_n}]
 \le (\E|H_n|^{2r})^{1/2}\Pp(D_n)^{1/2}
 \le C_r n^{3r/2}e^{-cn/2}.
\]
Indeed, $\|W_n\|_{2n}\le 2^{1/(2n)}\sqrt{n/2}$ in unit variance. On $D_n^c$, integration of $r t^{r-1}C'e^{-c't^2}$ bounds the $r$th moment uniformly. Thus $|H_n|^s$ is uniformly integrable after omission of finitely many initial dimensions, even when $s<1$. Apply \eqref{eq:main} first to $\min(|x|^s,M)$ and then send $M$ to infinity. The Gaussian moment in \eqref{eq:mom-main} equals
\[
 (\pi/2)^s\,\frac{2^{s/2}\Gamma((s+1)/2)}{\sqrt{\pi}} .
\]
Since $\E W_n=0$ by square-board symmetry, the case $s=2$ gives $n\,\sd(V_n)/\sigma\to\pi/2$ after restoring the scale. Fixing $s$ is essential even for bounded entries. For fair signs, a first row of ones has probability $2^{-n}$ and forces $V_n=1$. Hence $\E|nV_n|^n\ge(n/2)^n$, whereas the $n$th absolute Gaussian moment on the proposed scale is at most $(C\sqrt n)^n$.

\subsection{Support variance and a limitation of the deletion scores}
For symmetric atomless square games the finite-size inequality \eqref{eq:support-tail} controls the support without any entry moments. In particular, every fixed power of $(K-n/2)/\sqrt n$ has bounded absolute expectation uniformly in the dimension. For atomic input, the support comparison in Section~\ref{sec:fluctuations} gives a bounded difference from a smoothed support on good boards and a difference at most $n$ on the exponentially rare complement, with the smoothing coupling chosen at each size. Its mean squared difference divided by $n$ tends to zero. The smoothed supports satisfy \eqref{eq:support-tail} uniformly over the smoothing laws. The joint limit and this second-moment control give $\Var(K)/n\to1/8$ for every measurable attained support size.

This completes the proof of Theorem~\ref{thm:main}, together with the score limit in Section~\ref{sec:fluctuations} and Theorem~\ref{thm:paper-uniqueness}.

There is a limit to the likelihood argument even for atomless payoffs. For fixed $n\ge2$ and all $t>0$,
\[
 \E\exp\{t\sqrt n\,[(1-K/n)/F-1]\}=\infty .
\]
A pure saddle can be prescribed at a small quantile. For $0<u\le1/3$, take its entry between the $u$ and $2u$ quantiles, every other entry in its column below the lower quantile, and every other entry in its row above the upper one. This event has probability at least $u^{n^2}$, gives $K=1$ almost surely, and places $F$ in $[u,2u]$. As $u\downarrow0$, the exponential reward outgrows this polynomial probability cost. This explains the role of the two finite deletion likelihoods.

\section{Convex hulls, signs, and exact ties}\label{sec:ties}

The Gaussian law describes the value on a scale of order $1/n$. It leaves a finite-dimensional question open. How often does the value equal the symmetry center exactly? For this question we work with the centered matrix $B$, allow $R$ rows and $C$ columns, and assume only that its entries are independent copies of a symmetric random variable $Y$. Both dimensions are positive. Write $W=V(B)$. A tie means $W=0$. Matching pennies, for example, has a tie and a unique pair of optimal strategies.

\subsection{A convex feasibility identity}
Independently of $B$, retain each row with probability one half. Let $I$ be the set of retained rows and put
\[
 D_I=\min_{q\in\Delta_C}\max_{i\in I}|(Bq)_i|,\qquad D_\varnothing=0.
\]
Thus $D_I$ is the distance in the sup norm from the origin to the convex hull of the columns of $B$ projected onto $I$.

\begin{proposition}\label{prop:hull-paper}
One has $\max(W,0)\stackrel{\rm d}{=}D_I$. At the minimizing threshold, choose the feasible column strategy of smallest Euclidean norm on each side. The threshold and the chosen strategy have the same joint law. On $\{W\ge0\}$ the selection on the game side is an optimal strategy.
\end{proposition}
\begin{proof}
Fix $t\ge0$. For a convex set $F$ of candidate strategies and a linear form $f$, let $h(G)$ indicate that $G$ is nonempty. Then
\begin{align*}
 h(F\cap\{f\le t\})+h(F\cap\{-f\le t\})
   =h(F)+h(F\cap\{|f|\le t\}).
\end{align*}
Indeed, the two halfspaces cover $F$. If each meets $F$, the segment between witnesses meets their intersection. Applying this identity successively to the row forms on the simplex equates two counts. The first counts row sign choices for which some $q$ satisfies all oriented inequalities. The second counts the masks for which the retained absolute inequalities are feasible. The counts, divided by $2^R$ and averaged over $B$, are the two distribution functions at $t$, since the law of $B$ is invariant under each row sign change.

For the refinement, apply the same segment argument to pairs $(v,q)$ with threshold $v\ge0$, using the halfspaces $\pm(Bq)_i\le v$. Order feasible pairs first by $v$ and then by $\|q\|_2^2$. For fixed $(t,w)$ the set of pairs with $v<t$, together with the pairs with $v=t$ and $\|q\|_2^2<\|w\|_2^2$, is convex. It remains convex when $(t,w)$ is included, by strict convexity of squared norm on the boundary. The difference between the two hit indicators detects exactly the optimizer $(t,w)$. This identifies the multiplicity of every selected pair in the two finite lists. Compactness gives attainment, and the selection is measurable by finite-dimensional measurable minimization. Averaging gives the asserted joint law.
\end{proof}

For square games of finite positive variance, the value limit gives $\Pp(D_I>0)\to 1/2$. For $x>0$, divide $\Pp(nW_n/\sigma>x)$ by $\Pp(W_n>0)$ and apply Proposition~\ref{prop:hull-paper}. Conditional on the projected hull missing zero, $nD_I/\sigma$ converges in law to $(\pi/2)|Z|$ for a standard normal $Z$.

\subsection{Counting strict sign patterns}
Independently retain columns with probability one half as well, with retained set $J$. Write $i=|I|$, $j=|J|$, and $r=\operatorname{rank}B_{I,J}$, with rank zero for an empty side.
\begin{theorem}[Rank formula for a tie]\label{thm:tie-parity}
\[
 \Pp(W<0)=\E(-1)^{i-r},\qquad
 \Pp(W>0)=\E(-1)^{j-r}.
\]
Consequently
\begin{equation}\label{eq:tieparity}
 \Pp(W=0)=4\,\Pp\bigl(i\equiv j\bmod 2,\ 
                          \min(i,j)-r\ \text{is odd}\bigr).
\end{equation}
\end{theorem}
\begin{proof}
Fix a matrix. Its value is negative precisely when some $q\in\Delta_C$ satisfies $Bq<0$ coordinatewise. By strictness, $q$ can be perturbed into the interior of the simplex. Homogeneity then removes the normalization. Under row and column sign changes the negative-value sign configurations therefore correspond to the realizable strict sign patterns of the $R$ row forms and the $C$ coordinate forms on $\R^C$.

The region count for a real hyperplane arrangement \cite{Zas} gives the number of strict patterns of a finite list of linear forms as
$\sum_H(-1)^{|H|-\operatorname{rank}H}$, where $H$ runs through the sublists indexed by subsets. One proof adds forms successively. A new nonzero form splits precisely those open regions meeting its kernel. The additional count is the number of strict patterns of the preceding forms restricted to that kernel. The subset sum satisfies the same recursion, since ranks of sets containing the new form are one plus their restricted ranks. A zero form makes the two terms of every paired subset cancel, in agreement with the absence of strict patterns.

Choose the row forms indexed by $I$ and the coordinate forms indexed by $J^c$. Their rank is $C-j+r$, so their contribution is $(-1)^{i-r}$. Summation over masks and sign invariance prove the first identity. Applying it to $-B^\top$, whose value is $-W$, proves the second. Finally $\E(-1)^{i+j}=0$. For each mask,
\[
 1+(-1)^{i+j}-(-1)^{i-r}-(-1)^{j-r}
 =4\,{\bf1}_{\{i\equiv j\bmod2,\ \min(i,j)-r\ {\rm odd}\}} .
\]
Taking expectations proves \eqref{eq:tieparity}.
\end{proof}

For an atomless entry law all square minors have full rank almost surely, by conditioning on a cell with nonzero cofactor and inducting on the size. Thus the arrangement is in general position. Its strict region count \cite{Cover} is $2\sum_{k=0}^{C-1}\binom{R+C-1}{k}$, by the same restriction recursion. Division by $2^{R+C}$ gives the negative-value probability. For a general symmetric law one obtains
\begin{equation}\label{eq:cover-sandwich}
 \Pp(W>0)\ \le\ 
 \Pp\bigl(\operatorname{Bin}(R+C-1,1/2)\le R-1\bigr)
 \ \le\ \Pp(W\ge0).
\end{equation}
Indeed, adding independent symmetric atomless noise bounded by $\varepsilon$ to each cell changes the value by at most $\varepsilon$. Apply the atomless formula and let $\varepsilon$ decrease to zero. The gap between the endpoints of \eqref{eq:cover-sandwich} is exactly the tie probability.

There is already a useful moment-free square estimate. Set
$a=\max\{\Pp(Y=0),1/2\}<1$ when $Y$ is nondegenerate. Each point mass is at most $a$ by symmetry. A fresh vector with independent entries belongs to a fixed subspace of codimension $d$ with probability at most $a^d$, by exposing its nonpivot coordinates first. Exposing the shorter side of an $i$ by $j$ matrix vector by vector, and summing the probabilities of a first dependence, gives a deficiency bound $a^{|i-j|+1}/(1-a)$. For two independent $\operatorname{Bin}(n,1/2)$ masks, $\Pp(i-j=h)=4^{-n}\binom{2n}{n+h}\le1/\sqrt{n+1}$. Sum over $h$ in \eqref{eq:tieparity} to get
\[
 T_n:=\Pp(W_n=0)\ \le\ \frac{8a}{(1-a)^2\sqrt{n+1}} .
\]

\section{The exponential rate of an exact tie}\label{sec:tie-rates}

Let $S_n=\Pp(\det B_n=0)$, and let $Y'$ be an independent copy of $Y$. Define
\[
 p_0=\Pp(Y=0),\qquad q_2=\Pp(Y=Y'),\qquad
 \beta=\max\{p_0,q_2\},\qquad L(b)=\frac{(1+\sqrt b)^2}{4}.
\]
For $n\ge2$ a prescribed zero row and a prescribed pair of equal rows give $S_n\ge\beta^n$. The next result explains how these elementary singularity events are reflected in the game.

\begin{theorem}[Transfer in both directions]\label{thm:tie-transfer}
For every fixed nondegenerate symmetric law with at least one atom,
\begin{align}
 \liminf_{n\to\infty} T_n^{1/n}&\ge L(\beta),&
 \limsup_{n\to\infty} T_n^{1/n}
 &= L\bigl(\limsup_{n\to\infty}S_n^{1/n}\bigr).\label{eq:root-transfer}
\end{align}
In particular $T_n^{1/n}\to L(\beta)$ if and only if $S_n^{1/n}\to\beta$. No moment assumption is needed. For atomless laws \eqref{eq:tieparity} gives $T_n=0$ for every $n$.
\end{theorem}
\begin{proof}
First suppose $S_k\le D b^k$ for all positive $k$, with $0<b\le1$. Complete an $i$ by $j$ submatrix to a square of side $\max(i,j)$ with fresh entries. Deficiency forces singularity of the completion, because each appended row or column increases rank by at most one. Since $\max(i,j)\ge(i+j)/2$, \eqref{eq:tieparity} and two applications of the binomial theorem give
\begin{equation}\label{eq:rect-power-upper}
 \Pp(W=0)\le 4D\,\E\, b^{(i+j)/2}
       =4D\left(\frac{1+\sqrt b}{2}\right)^{R+C}.
\end{equation}
For squares this yields the upper bound in \eqref{eq:root-transfer} by decreasing $b$ to the upper singularity rate. At upper rate one the bound follows from $T_n\le1$.

For the reverse estimate let $O_k$ denote the probability of odd corank at size $k$. A positive even corank is at least two. Append a column and then a row. Each raises the rank with conditional probability at least $1-a$, by the subspace bound above. When both do so, the new corank is odd. Hence $S_k\le O_k+(1-a)^{-2}O_{k+1}$. At size $N$, the contribution to \eqref{eq:tieparity} from masks having $h$ rows and $h$ columns gives $4\binom Nh^2 O_h\le4^N T_N$. Use this at $N=n+d+1$ for $h=k+d$ and $k+d+1$. Both binomial coefficients are at least $\binom nk$. For a constant $D_1$ depending only on the law,
\begin{equation}\label{eq:one-mask}
 \binom nk^2 S_{k+d}\le D_1\,4^{n+d}T_{n+d+1}
 \qquad(0\le k\le n,\ d\ge0).
\end{equation}

By Cauchy--Schwarz and the binomial theorem,
\[
 \frac{L(b)^n}{n+1}\le
 4^{-n}\sum_{k=0}^n\binom nk^2 b^k\le L(b)^n .
\]
Insert $d=2$ into \eqref{eq:one-mask}, sum over $k$, and use $S_{k+2}\ge\beta^{k+2}$. This gives $\beta^2 L(\beta)^n\le16D_1(n+1)^2T_{n+3}$. Positive atomic mass ensures $\beta>0$, proving the lower-limit assertion.

It remains to recover the upper singularity rate even when it occurs along a sparse subsequence. Put $\tau=\limsup T_n^{1/n}$. For $x>2\sqrt{\tau}$ we have $x>1$ by the lower bound, and $T_n\le C(x^2/4)^n$. Use \eqref{eq:one-mask} with $d=0$ and take square roots to obtain $\binom nk\sqrt{S_k}\le C' x^n$. Multiply by $z^n$ and sum over $n\ge k$ for $0<z<1/x$. The identity $\sum_{n\ge k}\binom nk z^n=z^k/(1-z)^{k+1}$ gives
\[
 \sqrt{S_k}\le \frac{C'(1-z)}{1-xz}
                 \left(\frac{1-z}{z}\right)^k .
\]
Take $k$th roots first, then let $z\uparrow1/x$ and $x\downarrow2\sqrt\tau$. This gives $\limsup S_k^{1/k}\le(2\sqrt\tau-1)^2$, proving \eqref{eq:root-transfer}. Strict monotonicity of $L$ and $S_n\ge\beta^n$ give the equivalence.
\end{proof}

For symmetric laws of finite positive variance, the exponential singularity estimate proved in the formal development gives $S_k\le D b^k$ for some $b<1$. Equation~\eqref{eq:rect-power-upper} thus bounds ties by $C_0e^{-c_0(R+C)}$ uniformly in the aspect ratio, including a fixed shorter side.

\subsection{A sufficient condition on the atom masses}

Write $p(y)=\Pp(Y=y)$. The following criterion uses the amount of mass carried by points of very small individual probability.
\begin{proposition}\label{prop:small-mass}
Under the assumptions of Theorem~\ref{thm:tie-transfer}, if $n\,\Pp(p(Y)\le\beta^n)\to\infty$, then $S_n^{1/n}\to\beta$ and $T_n^{1/n}\to L(\beta)$.
\end{proposition}
\begin{proof}
Let $B_0$ be a measurable set of mass $q$ outside which each atom has mass at most $\eta$. Put $H_n(K)=\sum_{j=0}^K\binom nj$ for $1\le K\le n$. We claim
\begin{equation}\label{eq:light-atoms}
 S_n\le 2H_n(K)^2\beta^{n-K}+4^n q^{K^2}+n^2\eta .
\end{equation}
If some minimal row dependence has support $s\le K$, choose its rows and $s-1$ independent witness columns on those rows. The witness determines the coefficient vector up to scale. A fresh column must satisfy the corresponding homogeneous equation. For one nonzero coefficient the probability is $p_0$. With at least two, condition on all but two independent entries. The probability of the remaining affine equation is at most $q_2=\sum_y p(y)^2$ by Cauchy--Schwarz on the matched atom masses. Thus each fresh column passes with probability at most $\beta$. Enumerate the choices of rows and witnesses and allow the transpose to get the first term.

Otherwise choose a left and a right null vector. Both supports have more than $K$ indices. If all entries in their product rectangle lie in $B_0$, a union bound over at most $4^n$ rectangles gives the second term. If a cell $(i,j)$ there lies outside $B_0$, the left dependence forces row $i$ into the span of the other rows. The right null vector annihilates that span and has nonzero coordinate $j$, so $e_j$ is outside the span. After exposing all other cells, at most one value at $(i,j)$ can complete row $i$ into that span. Its probability outside $B_0$ is at most $\eta$. Summation over cells proves \eqref{eq:light-atoms}, including for corank larger than one.

Take $B_0=\{p>\beta^n\}$, $\eta=\beta^n$ and $K=\lfloor\delta n\rfloor$. For $0<\delta<1/2$, $H_n(K)\le e^{n h(\delta)}$ with $h(\delta)=-\delta\log\delta-(1-\delta)\log(1-\delta)$. The first term has upper root at most $e^{2h(\delta)}\beta^{1-\delta}$. The second has root tending to zero since $\log q\le-(1-q)$ and $n(1-q)\to\infty$. The last term has root limit $\beta$. Let $\delta\downarrow0$ and apply Theorem~\ref{thm:tie-transfer}.
\end{proof}

For a mixed law let $\alpha\in(0,1)$ be the total mass of its atoms. Take $B_0$ to be the set of all atoms and $\eta=0$ in \eqref{eq:light-atoms}. The bound becomes $2H_n(K)^2\beta^{n-K}+4^n\alpha^{K^2}$. The proof of this singularity bound uses independence alone. Even for an asymmetric mixed law, for every $b\in(\beta,1)$ there is $D_b$ such that $S_n\le D_b b^n$ for all $n$, and the matching lower bound yields $S_n^{1/n}\to\beta$. Symmetry enters when the rank formula transfers this result to the tie probability.

The small-atom condition can also hold for a bounded, purely atomic law on the rationals. Put probability $1/2$ at the origin. For every $k\ge0$, distribute mass $2^{-k-2}$ uniformly on $2^{(4^{k+1})^2}$ pairs of distinct signed rational radii. One explicit choice of positive radii is $1/(1+\langle k,j\rangle)$ for any injective pairing of natural numbers. Each atom has mass at most $1/2$, so $\beta=1/2$. If $4^k\le n<4^{k+1}$, each atom in layer $k$ has mass at most $2^{-n}$ while $n$ times the layer mass is at least $2^{k-2}$. The singularity and tie bases are therefore $1/2$ and $(3+2\sqrt2)/8$.

\subsection{Singularity estimates from the arithmetic of the support}

For fair signs, Tikhomirov's theorem gives $S_n^{1/n}\to1/2$ \cite{Tik}. Its Lean proof is included in the project. Theorem~\ref{thm:tie-transfer} gives $T_n^{1/n}\to(3+2\sqrt2)/8$. A second unconditional example follows from polynomial specialization. Fix a transcendental real number $t$ and set $Y=(1-t)\varepsilon+t\varepsilon'$ for independent fair signs. Its four atoms $\pm1,\pm(1-2t)$ have equal mass, so $\beta=1/4$. Realize the matrix as $(1-t)A+tB$ with independent fair-sign matrices $A,B$. Its determinant at variable $t$ is an integer polynomial. Vanishing at a transcendental point forces vanishing at both $0$ and $1$. Thus
\[
 4^{-n}\le S_n(Y)\le S_n(\varepsilon)^2 \qquad(n\ge2).
\]
The square on the right uses independence. It follows that the singularity base is $1/4$ and the tie base is $9/16$.

Two broader families use published singularity bounds as explicit inputs to the formal proofs.

\paragraph{Finite alphabets and algebraically independent radii.}
For any nonconstant finite real alphabet with its fixed cell law, Jain, Sah and Sawhney's least-singular-value bound at threshold zero gives, for each $\epsilon>0$ and all sufficiently large $n$,
\[
 S_n\le 2n p_0^n + (1+\epsilon)^n q_2^n
 \qquad\text{\cite[Theorem 1.4]{JSS}}.
\]
This proves the sharp bases $\beta$ and $L(\beta)$ for symmetric finite alphabets. It also applies to countable supports contained in $\{0,\pm a_0,\pm a_1,\ldots\}$ when the positive radii are algebraically independent over $\mathbb Q$. Specialize all but the first $K$ radii to zero. A singular determinant at the original radii is the zero polynomial, hence remains singular after specialization. For each fixed sufficiently large $K$ the finite-law upper bound applies. Under the natural cell coupling, the zero indicators stabilize almost surely as $K\to\infty$. The collision indicators stabilize for two independent coupled cells as well. The resulting finite-law bases therefore tend to $\beta$. Taking the dimension limit first proves the claim without uniformity in the finite-alphabet estimates. Algebraic independence is essential to this argument. Replacing $4$ by zero in the singular matrix with rows $(1,2)$ and $(2,4)$ produces determinant $-4$. For a countable example choose algebraically independent $a_j\in[1,2]$ and give each of $\pm a_j$ mass $2^{-j-2}$, $j\ge0$. Its collision base is $1/6$ and its small-atom tail is geometric, so specialization supplies the sharp rate beyond the scope of Proposition~\ref{prop:small-mass}.

\paragraph{Lattice laws with a nonnegative characteristic function.}
Suppose $Y$ is nondegenerate, symmetric, integer-valued, has finite variance and satisfies $\phi(t)=\E\cos(2\pi tY)\ge0$ everywhere. Fourier inversion gives $p(j)=\int_0^1\phi(t)\cos(2\pi jt)\,dt\le p_0$, hence $0<p_0<1$ and $\beta=p_0$. Choose $c$ with $e^{-2c}<p_0$. Moving entries with $|Y|>e^{cn}$ to zero changes the entire board with probability at most $n^2\E Y^2 e^{-2cn}$. For comparison, fix a cutoff $K$ retaining a nonzero atom and move only the mass outside $[-K,K]$ to zero. This fixed finite comparator has zero mass $a_K=p_0+\Pp(|Y|>K)<1$ and a positive minimum occupied mass. For $e^{cn}\ge K$, both truncations increase $\phi$ pointwise, and the larger increase occurs for the fixed comparator. Thus the modulus of the clipped input transform is bounded by the comparator transform. The clipped input has all atom probabilities at most $a_K$ and lies in an integer interval of size $e^{O(n)}=n^{o(n)}$. These are the rank-one, exponent-one hypotheses of Bourgain, Vu and Wood \cite[Definition 2.1 and Theorem 2.2]{BVW}. If strict positivity of the comparator transform is desired, mixing in $\epsilon$ mass at zero raises the ceiling to $(1-\epsilon)a_K+\epsilon$. Taking the dimension limit, then $\epsilon\downarrow0$ and $K\to\infty$, gives upper root $p_0$ for the original board. The zero-row bound gives equality. Thus the tie base is $L(p_0)$. Examples include symmetric integer laws with $p_0\ge1/2$ and nondegenerate differences $U-U'$ of independent copies of an integer-valued variable of finite variance, since their characteristic function is a squared modulus. Positive rescaling gives the same conclusion on $h\mathbb Z$.

\paragraph{An explicit boundary of the methods.}
Give each of $k,-k$ mass $2^{-k-1}$ for $k\ge1$. This law has finite variance, $p_0=0$, and $q_2=1/6$. For $u>0$ the total mass at atoms of individual weight at most $u$ is at most $4u$, by summing two geometric tails. Thus the small-atom criterion fails. The integer radii are algebraically dependent. The transform also takes negative values, since its integral is $p_0=0$ and $\phi(0)=1$.

There is a quantitative obstruction to improving the Fourier comparison by changing its exponent. For a fixed nondegenerate symmetric lattice law and every comparator with zero mass $p$ and $|\phi|^r\le\phi_{\rm comp}$ for an integer $r\ge1$, integration and Parseval's identity give
\[
 p^{1/r}\ge \|\phi\|_{L^r[0,1]}\ge
 \int_0^1|\phi(t)|\,dt>
 \int_0^1|\phi(t)|^2\,dt=q_2 .
\]
The strict inequality follows by continuity because $0<|\phi|<1$ on an interval of positive length. The gap is fixed by the input law, uniformly over all these comparators and exponents. For the signed geometric law, \eqref{eq:root-transfer} gives the lower tie base $L(1/6)$. Whether the matching upper bound holds for this law remains open in the present argument.

\section{Changing the shape or the law of the board}\label{sec:extensions}
The square-board theorem uses both the equality of the two dimensions and the fact that the payoff law is fixed. Varying these assumptions separately explains which parts of the limit are stable.

\subsection{Rectangles in the critical window}
Suppose $r_n=n+a\sqrt n+o(\sqrt n)$ and $c_n=n+b\sqrt n+o(\sqrt n)$. Let $W=V(A_{r_n,c_n})-m$. For a fixed symmetric atomless law of finite positive variance the common support size $K$ and value satisfy
\begin{equation}\label{eq:critical-rectangle}
 \left(\frac{K-n/2}{\sqrt n},\frac{nW}{\sigma}\right)
 \Longrightarrow
 \left(\frac{a+b}{4}+\frac{Z_1}{\sqrt8},\
       \frac{\pi(a-b)}{2\sqrt2}+\frac{\pi}{2}Z_2\right),
\end{equation}
with independent $Z_1,Z_2$. For arbitrary atomic mass, the value marginal in \eqref{eq:critical-rectangle} and $\Pp(W>0)\to\Phi((a-b)/\sqrt2)$ hold on the original rectangle, where $\Phi$ is the standard normal distribution function.

Indeed, on scores with limiting independent coordinates $u_1,u_2$ of variance $1/4$, the likelihood for the displaced dimensions is asymptotically
\[
 \exp\{2a u_1+2b u_2-(a^2+b^2)/2\}.
\]
Completing the square shifts the means by $(a/2,b/2)$ and preserves the covariance. The half-sum and the difference then give \eqref{eq:critical-rectangle}. The finite likelihood convergence uses truncation outside compact score sets. A proof for panels obtained by adjoining rows or columns suffices: start from the smaller dimension and recenter the support by the deterministic difference of centers. For atomic entries use bounded smoothing of radius $\min(r_n,c_n)^{-2}$. This gives an $o(1)$ error in $nW$ by the value Lipschitz bound. The Gaussian limit assigns no mass to zero, so its strict winning probability is the limit on the original boards.

\subsection{Changing the payoff law}
Let $Y_n$ have centered symmetric atomless law $\nu_n$, and suppose there exist couplings with a fixed symmetric $Y$ of variance $\sigma^2\in(0,\infty)$ for which $\E|Y_n-Y|^2\to0$. The reference $Y$ may have atoms. The joint square-board support and value limit, both variance limits, and all fixed positive absolute value moments hold with the \emph{reference} scale $\sigma$. The coupling concerns one cell. It need not make the largest of the $n^2$ entry errors small on the value scale.

The transfer uses the coupling at the level of directions and fresh projections. For independent coupled cells and every deterministic unit vector $x$,
\[
 \E\left|\sum_j x_j(Y_{n,j}-Y_j)\right|^2=\E|Y_n-Y|^2 .
\]
Symmetry centers each difference. The corresponding direction estimates for feasible strategies, together with the local estimates for their projected sums, give the fixed-board score limit along the array. The support estimate \eqref{eq:support-tail} is uniform over all the atomless laws. The value argument of Section~\ref{sec:moments} applies with uniform moderate-deviation constants along the sequence and a common second-moment bound.

Equal variance alone allows a different limit. Take entries equal to $+n$ and $-n$ with probability $1/(2n^2)$ each and zero otherwise. Every entry has variance one. A specified row and column are both zero except on an event of probability at most $(2n-1)/n^2$. On the former event their pure strategies guarantee zero to both players. Thus $\Pp(V_n\ne0)\le 2/n$ and $nV_n\to0$ in probability. This example has atomic input laws.

\subsection{A single smooth law can converge arbitrarily slowly}
Fix $s>0$ and a positive sequence $\rho_n\to0$. There exists a symmetric variance-one entry law with a strictly positive $C^\infty$ density and with infinite $(2+\delta)$th absolute moment for every $\delta>0$, for which each of the following errors divided by $\rho_n$ tends to infinity along the same increasing subsequence:
\[
 \left|\E|nV_n|^s-M_s\right|,\quad
 \left|\Pp(nV_n\le\pi/2)-\Phi(1)\right|,\quad
 \sup_x\left|\Pp(nV_n\le x)-\Phi(2x/\pi)\right|
 .
\]
Here $M_s=(\pi/2)^s\E|Z|^s$ for a standard normal $Z$. The quantifiers allow the law to depend on $s$ and $\rho$.

To construct it, distribute a fixed variance budget among a sequence of increasingly rare symmetric pairs of spikes, and convolve with a fixed nondegenerate Gaussian. After the first $k$ layers, the variance is $\sigma_k^2<1$, increasing to one. If the remaining variance is placed sufficiently far in the tails, the probability that a board of a chosen size $n_k$ sees any later layer can be made as small as required. The truncated law has limiting $s$th moment $\sigma_k^s M_s$ and limiting distribution function $\Phi(1/\sigma_k)$ at $\pi/2$. Both are separated from their targets by a positive gap at this stage. Choose $n_k$ so large that the truncated law already realizes both gaps, and $k\rho_{n_k}$ is smaller than a fixed fraction of the smaller gap.

The moment comparison requires more than a bound on the probability of observing a later spike. Couple a completed law to its truncation with one-cell disagreement probability $\delta_k$. For $s\le n$ and a cutoff $D>0$, clipping $|nV_n|^s$ at $D$ gives the two-law moment error bound
\[
 2n^2\delta_kD+6n^{3s}/D.
\]
The first term is the board disagreement cost. The second follows from the $2s$-moment estimate in \eqref{eq:critical-integrability}. First choose $D$ large for the already selected $n_k$, then choose all future layer probabilities sufficiently small. The same coupling bounds the distribution-function error by $n_k^2\delta_k$. Future layers can meet every preceding budget while retaining their prescribed total variance by increasing their amplitudes. Choosing those amplitudes to grow sufficiently fast gives divergence of every higher entry moment. The Gaussian convolution supplies the positive smooth density. This diagonal choice produces a single completed law and a common subsequence.

\section{What the argument uses from the game}\label{sec:portable}
Three deductions above apply to other random feasible sets. Their assumptions describe what must be checked in a new model.

\paragraph{Counting and conditioning.}
Suppose each feasible optimum can be counted once by a marked set of active constraints, and new constraints of type $j$ pass the marked point independently with probability $f_j$. Independent Poisson counts then give conditionally independent unused counts of means $\lambda_j f_j$ by the factorial identity in Section~\ref{sec:fluctuations}. For a phase $w$ with $|w|\le1$ and a conditional noise transform $k$, the elementary bound
\[
 \left|\E(wk)-q\,\E w\right|\le \E|k-q|
\]
allows division by a limiting factor $q\ne0$. To return to fixed dimensions one must additionally control the local probability of the desired counts. When one optimizer has several representing bases, a counting convention or multiplicity correction is needed before the Poisson calculation.

\paragraph{Averaging signs.}
For a convex feasible set and a nonnegative threshold, the number of feasible systems obtained by orienting each linear inequality equals the number obtained by imposing its two-sided slab or omitting it. The segment argument in Section~\ref{sec:ties} gives the identity one constraint at a time. For example, take $k$ distinct points in an interval and at each point choose the left or the right half-interval. In the equality version only the empty choice and the $k$ singletons are feasible, so precisely $k+1$ of the $2^k$ orientations are feasible. Convexity matters. On $\{-1,1\}$ the two orientations of $x\le0$ are both feasible, whereas of the options $x=0$ and omission only omission is feasible.

\paragraph{Testing an independent constraint.}
Suppose removal of constraint $i$ leaves data $D_i$ independent of the removed observation $X_i$. If failure outside an exceptional set $E_0$ entails a hit of some candidate $j$ in a finite set $J_i(D_i)$, and $b_{ij}(D_i)$ bounds the conditional probability of that hit, then
\[
 \Pp(\mathrm{failure})\le \Pp(E_0)+
        \sum_i \E\sum_{j\in J_i(D_i)} b_{ij}(D_i).
\]
This is conditional integration followed by a union bound. A pool selected after seeing $X_i$ could contain its exact realization even if each predetermined candidate had negligible hit probability. In a game the geometric work identifies a finite pool, detects failure and bounds its size. Those are the steps to establish in another optimization problem.

The limiting distribution still depends on the model. For the program $\max x$ subject to $x\le U_i$, with $n$ independent uniform variables on $[0,1]$, the optimum is $\min_i U_i$. Its rescaled survival function is $(1-z/n)^n$ for $0\le z\le n$, giving an exponential limit. The two Gaussian variables in a matrix game arise from its two equilibrium tests and the local geometry of their projections.

\appendix
\section{Formal statements and reproducibility}\label{sec:formal}
The proofs cited in the text belong to the two Lean projects in \cite{Project}. Their endpoint declarations retain the mathematical hypotheses. In particular, the reductions using the finite-alphabet theorem of Jain, Sah and Sawhney and the exponent-one estimate of Bourgain, Vu and Wood take those estimates as explicit premises. The fair-sign proof includes the Bernoulli singularity theorem from its bundled Lean source. The square limit, the geometric identities, the tie--singularity equivalence, and the small-atom criterion are independent of both external premises.

The Lean square dimension is often written $k+1$, which covers all positive sizes. Support is counted by strictly positive weights. The main limit quantifies over measurable \emph{attained sizes}, a formulation that covers measurable selections of probability vectors and also the smallest or largest attained size. Centering is applied to the matrix before singularity is tested, and both support choices are evaluated on the same original realization. The following links give the main formal statements.
\begin{itemize}
\item \href{https://github.com/Sheltercosmo/minimax-universality/blob/557c5e4321a3d9d2dcf969999462e8f380f8aa05/proofs/matrix-games/Extensions/FluctuationsAboutTheSymmetryCenter.lean}{Joint law and exact tie on the original board} and \href{https://github.com/Sheltercosmo/minimax-universality/blob/557c5e4321a3d9d2dcf969999462e8f380f8aa05/proofs/matrix-games/Extensions/RandomGameUniversality.lean}{joint limit, moments and complementary conclusions}.
\item \href{https://github.com/Sheltercosmo/minimax-universality/blob/557c5e4321a3d9d2dcf969999462e8f380f8aa05/proofs/matrix-games/MatrixGames/ExponentialStrategyUniqueness.lean}{Exact uniqueness and equal support} and \href{https://github.com/Sheltercosmo/minimax-universality/blob/557c5e4321a3d9d2dcf969999462e8f380f8aa05/proofs/matrix-games/MatrixGames/OptimalWeightTransport.lean}{uniform quadratic transport for optimal weights}.
\item \href{https://github.com/Sheltercosmo/minimax-universality/blob/557c5e4321a3d9d2dcf969999462e8f380f8aa05/proofs/matrix-games/MatrixGames/RectangularTieRankFormula.lean}{Finite rank identity} and \href{https://github.com/Sheltercosmo/minimax-universality/blob/557c5e4321a3d9d2dcf969999462e8f380f8aa05/proofs/matrix-games/Extensions/TieAndSingularityRateEquivalence.lean}{necessary and sufficient condition for the sharp tie base}.
\item \href{https://github.com/Sheltercosmo/minimax-universality/blob/557c5e4321a3d9d2dcf969999462e8f380f8aa05/proofs/matrix-games/Extensions/FiniteAlphabetSingularityRates.lean}{Finite-alphabet premise and deduction} and \href{https://github.com/Sheltercosmo/minimax-universality/blob/557c5e4321a3d9d2dcf969999462e8f380f8aa05/proofs/matrix-games/Extensions/LatticeTieAndSingularityRates.lean}{lattice premise and deduction}.
\end{itemize}
The general project records 3,300 theorem declarations replayed by the Lean 4.19.0 kernel. The Bernoulli project records 3,123 under Lean 4.33.0. These inventories include private and generated declarations and overlap in mathematical content. The transitive axiom audits admit \texttt{propext}, \texttt{Classical.choice}, and \texttt{Quot.sound}. The \href{https://github.com/Sheltercosmo/minimax-universality/blob/557c5e4321a3d9d2dcf969999462e8f380f8aa05/verification/proof-verification.md}{verification record} provides the build logs, kernel replay logs, and theorem inventories. To reproduce both builds from this revision, run \texttt{python verify.py --fetch} with Elan's Lake launcher available. The faster \texttt{python verify.py --source-only} checks release integrity and exact finite examples.

\subsection*{Proofs by section}
\noindent\textit{Section~\ref{sec:main}.} \href{https://github.com/Sheltercosmo/minimax-universality/blob/557c5e4321a3d9d2dcf969999462e8f380f8aa05/proofs/matrix-games/MatrixGames/SubgaussianGameFluctuations.lean}{The two continuous specializations}.

\noindent\textit{Section~\ref{sec:fluctuations}.} \href{https://github.com/Sheltercosmo/minimax-universality/blob/557c5e4321a3d9d2dcf969999462e8f380f8aa05/proofs/matrix-games/MatrixGames/MarkedOptimalSupport.lean}{The finite marked-support identity}, \href{https://github.com/Sheltercosmo/minimax-universality/blob/557c5e4321a3d9d2dcf969999462e8f380f8aa05/proofs/matrix-games/MatrixGames/SupportConcentration.lean}{Finite-size support concentration}, \href{https://github.com/Sheltercosmo/minimax-universality/blob/557c5e4321a3d9d2dcf969999462e8f380f8aa05/proofs/matrix-games/Extensions/MarkedPoissonCounts.lean}{The finite-type identity}, \href{https://github.com/Sheltercosmo/minimax-universality/blob/557c5e4321a3d9d2dcf969999462e8f380f8aa05/proofs/matrix-games/MatrixGames/FixedBoardScoreCLT.lean}{The two score limits}, \href{https://github.com/Sheltercosmo/minimax-universality/blob/557c5e4321a3d9d2dcf969999462e8f380f8aa05/proofs/matrix-games/Extensions/RandomConstraintLemmas.lean}{Conditional characteristic-function division}, \href{https://github.com/Sheltercosmo/minimax-universality/blob/557c5e4321a3d9d2dcf969999462e8f380f8aa05/proofs/matrix-games/MatrixGames/PoissonLikelihoodUntilting.lean}{The conditioning estimate in \(L^1\)}, \href{https://github.com/Sheltercosmo/minimax-universality/blob/557c5e4321a3d9d2dcf969999462e8f380f8aa05/proofs/matrix-games/MatrixGames/ResidualMatrixNormBounds.lean}{The finite-variance residual estimate}, \href{https://github.com/Sheltercosmo/minimax-universality/blob/557c5e4321a3d9d2dcf969999462e8f380f8aa05/proofs/matrix-games/MatrixGames/WinningStrategyShape.lean}{Recovery from the winning side}, \href{https://github.com/Sheltercosmo/minimax-universality/blob/557c5e4321a3d9d2dcf969999462e8f380f8aa05/proofs/matrix-games/MatrixGames/JointLimitFromStrategyShape.lean}{The inversion on the original square}, \href{https://github.com/Sheltercosmo/minimax-universality/blob/557c5e4321a3d9d2dcf969999462e8f380f8aa05/proofs/matrix-games/MatrixGames/JointLimitsForAllAtomicMasses.lean}{The fixed-surplus theorem}, \href{https://github.com/Sheltercosmo/minimax-universality/blob/557c5e4321a3d9d2dcf969999462e8f380f8aa05/proofs/matrix-games/MatrixGames/PerturbedSupportEnvelope.lean}{The support interval}, \href{https://github.com/Sheltercosmo/minimax-universality/blob/557c5e4321a3d9d2dcf969999462e8f380f8aa05/proofs/matrix-games/MatrixGames/AtomicSupportTransfer.lean}{Transfer to measurable attained sizes}, \href{https://github.com/Sheltercosmo/minimax-universality/blob/557c5e4321a3d9d2dcf969999462e8f380f8aa05/proofs/matrix-games/MatrixGames/OptimalWeightTransport.lean}{Quadratic transport for all optimizers}, \href{https://github.com/Sheltercosmo/minimax-universality/blob/557c5e4321a3d9d2dcf969999462e8f380f8aa05/proofs/matrix-games/MatrixGames/PositiveWeightsAndHullDistance.lean}{The law on positive coordinates}.

\noindent\textit{Section~\ref{sec:unique}.} \href{https://github.com/Sheltercosmo/minimax-universality/blob/557c5e4321a3d9d2dcf969999462e8f380f8aa05/proofs/matrix-games/MatrixGames/SupportPerturbationExample.lean}{The two support sets}, \href{https://github.com/Sheltercosmo/minimax-universality/blob/557c5e4321a3d9d2dcf969999462e8f380f8aa05/proofs/matrix-games/MatrixGames/OptimalVertexLatticeBounds.lean}{Point probabilities at optimal vertices}, \href{https://github.com/Sheltercosmo/minimax-universality/blob/557c5e4321a3d9d2dcf969999462e8f380f8aa05/proofs/matrix-games/MatrixGames/RowDeletedCandidatePools.lean}{Measurable lists from deleted data}, \href{https://github.com/Sheltercosmo/minimax-universality/blob/557c5e4321a3d9d2dcf969999462e8f380f8aa05/proofs/matrix-games/MatrixGames/CandidatePoolCoverage.lean}{probability of coverage}, \href{https://github.com/Sheltercosmo/minimax-universality/blob/557c5e4321a3d9d2dcf969999462e8f380f8aa05/proofs/matrix-games/MatrixGames/ObservableNonuniquenessBound.lean}{Conditional hit estimate}, \href{https://github.com/Sheltercosmo/minimax-universality/blob/557c5e4321a3d9d2dcf969999462e8f380f8aa05/proofs/matrix-games/MatrixGames/MixedLawUniquenessAndScores.lean}{Positive diffuse mass}, \href{https://github.com/Sheltercosmo/minimax-universality/blob/557c5e4321a3d9d2dcf969999462e8f380f8aa05/proofs/matrix-games/MatrixGames/FiniteGameLaws.lean}{Finite-game equivalences and bounds}, \href{https://github.com/Sheltercosmo/minimax-universality/blob/557c5e4321a3d9d2dcf969999462e8f380f8aa05/proofs/matrix-games/MatrixGames/ZeroRowSupportWidthWitness.lean}{The maximal support gap}.

\noindent\textit{Section~\ref{sec:moments}.} \href{https://github.com/Sheltercosmo/minimax-universality/blob/557c5e4321a3d9d2dcf969999462e8f380f8aa05/proofs/matrix-games/MatrixGames/SharpValueMomentCounterexample.lean}{Finite-variance witness}, \href{https://github.com/Sheltercosmo/minimax-universality/blob/557c5e4321a3d9d2dcf969999462e8f380f8aa05/proofs/matrix-games/MatrixGames/ModerateThresholdSlopes.lean}{Uniform increments on the moderate range}, \href{https://github.com/Sheltercosmo/minimax-universality/blob/557c5e4321a3d9d2dcf969999462e8f380f8aa05/proofs/matrix-games/MatrixGames/FiniteVarianceValueTails.lean}{Value tails on the original board}, \href{https://github.com/Sheltercosmo/minimax-universality/blob/557c5e4321a3d9d2dcf969999462e8f380f8aa05/proofs/matrix-games/MatrixGames/SupportSelectionVariance.lean}{Variance transfer for support selections}, \href{https://github.com/Sheltercosmo/minimax-universality/blob/557c5e4321a3d9d2dcf969999462e8f380f8aa05/proofs/matrix-games/MatrixGames/DeletionScoreMomentDivergence.lean}{Divergence for a fixed board size}.

\noindent\textit{Section~\ref{sec:ties}.} \href{https://github.com/Sheltercosmo/minimax-universality/blob/557c5e4321a3d9d2dcf969999462e8f380f8aa05/proofs/matrix-games/MatrixGames/LeastEnergyStrategyLaw.lean}{The selection and its measurability}, \href{https://github.com/Sheltercosmo/minimax-universality/blob/557c5e4321a3d9d2dcf969999462e8f380f8aa05/proofs/matrix-games/MatrixGames/RectangularTieRankFormula.lean}{The finite sign counts}.

\noindent\textit{Section~\ref{sec:tie-rates}.} \href{https://github.com/Sheltercosmo/minimax-universality/blob/557c5e4321a3d9d2dcf969999462e8f380f8aa05/proofs/matrix-games/MatrixGames/InverseBinomialTieTransform.lean}{Inversion of the binomial transform}, \href{https://github.com/Sheltercosmo/minimax-universality/blob/557c5e4321a3d9d2dcf969999462e8f380f8aa05/proofs/matrix-games/Extensions/TieAndSingularityRateEquivalence.lean}{The sharp rate equivalence}, \href{https://github.com/Sheltercosmo/minimax-universality/blob/557c5e4321a3d9d2dcf969999462e8f380f8aa05/proofs/matrix-games/Extensions/SmallAtomSingularityBound.lean}{The finite atom bound}, \href{https://github.com/Sheltercosmo/minimax-universality/blob/557c5e4321a3d9d2dcf969999462e8f380f8aa05/proofs/matrix-games/Extensions/SmallAtomRateCriterion.lean}{The limiting criterion}, \href{https://github.com/Sheltercosmo/minimax-universality/blob/557c5e4321a3d9d2dcf969999462e8f380f8aa05/proofs/matrix-games/MatrixGames/MixedLawSingularityAndTieRates.lean}{The mixed-law estimates}, \href{https://github.com/Sheltercosmo/minimax-universality/blob/557c5e4321a3d9d2dcf969999462e8f380f8aa05/proofs/matrix-games/Extensions/RationalAtomicTieRate.lean}{Construction of the rational payoff law}, \href{https://github.com/Sheltercosmo/minimax-universality/blob/557c5e4321a3d9d2dcf969999462e8f380f8aa05/proofs/bernoulli-tie-rates/RademacherTieRate.lean}{Fair signs}, \href{https://github.com/Sheltercosmo/minimax-universality/blob/557c5e4321a3d9d2dcf969999462e8f380f8aa05/proofs/bernoulli-tie-rates/FourPointTieAndSingularityRates.lean}{The four-point law}, \href{https://github.com/Sheltercosmo/minimax-universality/blob/557c5e4321a3d9d2dcf969999462e8f380f8aa05/proofs/matrix-games/Extensions/FiniteAlphabetSingularityRates.lean}{The external finite-alphabet estimate and its consequences}, \href{https://github.com/Sheltercosmo/minimax-universality/blob/557c5e4321a3d9d2dcf969999462e8f380f8aa05/proofs/matrix-games/Extensions/LatticeTieAndSingularityRates.lean}{The exponent-one premise and the lattice deduction}, \href{https://github.com/Sheltercosmo/minimax-universality/blob/557c5e4321a3d9d2dcf969999462e8f380f8aa05/proofs/matrix-games/Extensions/FourierComparisonGap.lean}{Uniform gap for Fourier comparators}, \href{https://github.com/Sheltercosmo/minimax-universality/blob/557c5e4321a3d9d2dcf969999462e8f380f8aa05/proofs/matrix-games/Extensions/SignedGeometricAtomTail.lean}{The geometric atom tail}.

\noindent\textit{Section~\ref{sec:extensions}.} \href{https://github.com/Sheltercosmo/minimax-universality/blob/557c5e4321a3d9d2dcf969999462e8f380f8aa05/proofs/matrix-games/MatrixGames/CriticalRectangularFluctuations.lean}{Rectangular joint law}, \href{https://github.com/Sheltercosmo/minimax-universality/blob/557c5e4321a3d9d2dcf969999462e8f380f8aa05/proofs/matrix-games/MatrixGames/UnperturbedRectangularValueLimit.lean}{Original-board value and sign}, \href{https://github.com/Sheltercosmo/minimax-universality/blob/557c5e4321a3d9d2dcf969999462e8f380f8aa05/proofs/matrix-games/MatrixGames/TriangularUniversality.lean}{Triangular joint theorem}, \href{https://github.com/Sheltercosmo/minimax-universality/blob/557c5e4321a3d9d2dcf969999462e8f380f8aa05/proofs/matrix-games/MatrixGames/TriangularValueMoments.lean}{Triangular moment estimates}, \href{https://github.com/Sheltercosmo/minimax-universality/blob/557c5e4321a3d9d2dcf969999462e8f380f8aa05/proofs/matrix-games/MatrixGames/SparseTriangularCounterexample.lean}{A variance-one sparse array}, \href{https://github.com/Sheltercosmo/minimax-universality/blob/557c5e4321a3d9d2dcf969999462e8f380f8aa05/proofs/matrix-games/MatrixGames/CoupledValueMomentBounds.lean}{Truncation estimate}, \href{https://github.com/Sheltercosmo/minimax-universality/blob/557c5e4321a3d9d2dcf969999462e8f380f8aa05/proofs/matrix-games/MatrixGames/ArbitrarilySlowMomentConvergence.lean}{Simultaneous lower bounds on the errors}.

\noindent\textit{Section~\ref{sec:portable}.} \href{https://github.com/Sheltercosmo/minimax-universality/blob/557c5e4321a3d9d2dcf969999462e8f380f8aa05/proofs/matrix-games/Extensions/MarkedPoissonCounts.lean}{Finite-type factorization}, \href{https://github.com/Sheltercosmo/minimax-universality/blob/557c5e4321a3d9d2dcf969999462e8f380f8aa05/proofs/matrix-games/Extensions/RandomConstraintLemmas.lean}{Transform and counting lemmas}, \href{https://github.com/Sheltercosmo/minimax-universality/blob/557c5e4321a3d9d2dcf969999462e8f380f8aa05/proofs/matrix-games/Extensions/RandomConstraintExamples.lean}{Interval and optimization examples}.

\end{document}